\documentstyle{article}

\def\QQa{\renewcommand{\baselinestretch}{1.3}\Huge\normalsize\large\small}

\begin{document}
\QQa

\large
\begin{flushright}
ITP.SB-93-06\\
Jan 10,1993
\end{flushright}

\begin{center}
\huge
Quantum Double of Heisenberg-Weyl Algebra,its Universal R-Matrix and their
Representations\\
\vspace{1cm}
\Large

Chang-Pu Sun \footnote {\large
Permanet address:Physics Dpartment,Northeast Normal University,
Changchun 130024,P.R.China}\\

Institute for Theoretical Physics,State University of New York,Stony Brook,
NY 11794-3840,USA\\
M.L.Ge \footnote {\large
Permanet address:Theoretical Physics Division,Nankai Institute of Methematics
Tianjin 30071,P.R.China}\\

Department of Physics,The University of Utah,Salt Lake City,
Utah 84112,USA\\
\vspace{1cm}
\huge
Abstract\\
\end{center}
\large
In this paper a new quasi-triangular Hopf algebra as the quantum
double of the Heisenberg-Weyl algebra is presented.Its universal R-matrix is
built and the corresponding representation theory are studied with the explict
construction for the representations of this quantum double.
\newpage

{\bf 1.Introduction}
\vspace{0.5cm}

Recently the quantum group theory associated with Yang-Baxter equation
 for nonlinear integrable system has became a focus of the attention from
both theoretical physicists and mathematicians [1].As a kernal
of this theory,Drinfeld's quantum double construction is a quite powerful
tool in constructing the solutions,namely the R-matrices,for quantum
Yang-Baxter equation (QYBE) in connection with certain algebraic structures,
such as quantum algebras[2],quantum superalgebras[3],quantum affine algebras[2]
,their multiparameter deformations[4]
and  the quantum doubles of the Borel subalgebras for universal enveloping
algebras(UEAs) of classical Lie algebra[6]

In this paper we will present a different quasi-triangular Hopf algebra
that is the quantum double of the Heisenberg-Weyl(HW) algebra based on
Drinfeld's quantum double construction.To construct the explicit
R-matrices for QYBE  from its universal
R-matrix  of this quantum double,we study its representation theory
and explicitly construct its finite and infinite dimensional representations.
A 6-dimensional example of R-matrices for this quantum double is given
as an illusration.The studies in this paper shows that ,like its q-deformation
[6-8],the ordinary HW algebra also realizes a so-called `quantum
group structure',quasi-triangular Hopf algebra associated QYBE.This fact
shoows that the canonical quantization defined by the HW algebra possibly
prompts the important role of `quantum group structure' and the QYBE in
quantum theory.
\vspace{0.6cm}

{\bf 2.Quantum Double of HW Algebra}
\vspace{0.4cm}

The Heisenberg-Weyl algebra (HW) algebra {\bf A} is an associative algebra
 generated by $a,\bar{a},E $ and the unit 1 .These generators
satisfy the defining relations
$$[a,\bar{a}]=E,[E,a]=0=[E,\bar{a}],\eqno{(2.1)}$$
If we take a special representation {\bf T} such that
$${\bf T}(a)^+={\bf T}(\bar{a}),{\bf T}(E)=unit~~ matrix~I$$
then $\bar{a}$ and $a$ can be regarded the creation and annihilation operators
of boson states in second quatization.Since the algebra {\bf A} is the UEA of
the HW Lie algebra with basis $\{a,{\bar a},E\}$,{\bf A} can be endowed with
a well-known Hopf algebraic structure
$$\Delta(x)=x\otimes 1+1\otimes x,S(x)=-x,\epsilon(x)=0,\eqno{(2.2)}$$
for $x=a,{\bar a},E$ where the algebraic homomorphisms $\Delta,\epsilon$ and
the algebraic antihomomorphism $S$ defined only for the generators
are naturally extended for the whole
algebra.According to the PBW theorem,the basis for the algebra  {\bf A} is
chosen as
$$\{X(m,n,s)=\bar{a}^ma^nE^s,m,n,s,\in {\bf Z }^+=\{0,1,2,..\}\}$$

Now,let us consider the dual Hopf algebra {\bf B} of {\bf A}.Suppose
$\bar b  ,b$ and $H$ are the dual generators to  $\bar a,a$ and $E$
respectively and then defined by
$$<X(m,n,s),\bar b>=\delta_{m,1}\delta_{n,0}\delta_{s,0}$$
$$<X(m,n,s), b>=\delta_{m,0}\delta_{n,1}\delta_{s,0} \eqno{(2.3)}$$
$$<X(m,n,s),E >=\delta_{m,0}\delta_{n,0}\delta_{s,1}$$
Since the algebra {\bf A} is commutative,its Hopf algebraic dual is
Abelian,ie.,the dual generators commute each other.Chosing a basis for {\bf B}
$$Y(m,n,s)=(m!n!s!)^{-1}\bar{b}^mb^nH^s,m,n,s,\in {\bf Z}^+,\eqno{(2.4)} $$
we prove the following proposition
\vspace{0.4cm}

{\bf Proposition 1}.{\it The equations (2.4) define a dual basis Y(m,n,s)
for  {\bf B} satisfying}
$$<X(m,n,s),Y(m',n',s')>=\delta_{m,m'}\delta_{n,n'}\delta_{s,s'},\eqno{(2.5)}$$

{\bf Proof}.According to the Hopf algebraic duality beween  {\bf A} and
   {\bf B}:
$$<a, b_1 b_2>=<\Delta_A (a), b_1\otimes b_2>, a\in {\bf A},~b_1,~b_2\in
{\bf B},$$
$$<a_1 a_2,b>=<a_2\otimes a_1,\Delta_B(b)>,a_1,a_2\in  {\bf A}
,b\in  {\bf B}$$
$$<1_A,b>=\epsilon_B(b),b\in {\bf B},\eqno{(2.6)}$$
$$<a,1_B>=\epsilon_A(a),a\in  {\bf A}$$
$$<S_A(a),S_B(b)>=<a,b>,a\in  {\bf A},b\in{\bf B} $$
where for C={\bf A,B},$\Delta_C,\epsilon_C$ and $S_C $ are
the coproduct,counit and
antipode of C respectively;$1_C$ is the unit of C.Without confusion we no
longer use the index C to specify,$\Delta_C,\epsilon_C$ and $S_C $
Let $G=\bar a,a  E$ corresponding to
 $F=\bar b,b,E$ respectively. For $G \neq \bar a$,
$$<\bar a^s G^m,F^n>=<\Delta(\bar a^s G^m),F^{n-1}\otimes F>$$
$$=\sum _{k,l=0}^{\infty}\frac{m!s!}{(m-k)!k!(s-l)!l!}
<\bar a^{s-l} G^{m-k}\otimes\bar a^l G^k  ,F^{n-1}\otimes F>$$
$$=m<\bar a^s G^{m-1},F^{n-1}>
=m!\delta_{m,n}\delta{s,0}$$

For $F\neq \bar b$,similarly,we have
$$<G^m,\bar b^sF^n>=m!\delta_{m,n}\delta{s,0}$$
$$<G^m,F^n>=m!\delta_{m,n}$$

Then,we have
$$<\bar a^m a^n,\bar b^s b^l>=<\Delta(\bar a^ma^n),\bar b^s\otimes b^l>
=s!l!\delta_{m,s}\delta_{n,l}$$
and therefore prove eq.(2.5).

It follows from the above proposition that
$$<X(m,n,s)\otimes X(k,l,r),\Delta(H)>=<X(k,l,r)X(m,n,s),H>$$
$$=\delta_{m,0}\delta_{n,0}\delta_{k,0}\delta_{l,0}(\delta_{s,1}
\delta_{r,0}+\delta_{r,1}\delta_{s,0})
+\delta_{m,1}\delta_{n,0}\delta_{s,0}\delta_{k,0}\delta_{l,1}\delta_{r,0}$$
namely,
$$\Delta(H)=H\otimes 1+1\otimes H +\bar b \otimes b$$
Similarly,we calculate other operations of the generators for {\bf B} under
$\Delta,\epsilon$ and $S$.The results are summarized as follows.

{\bf Proposition 2.}.{\it The dual Hopf algebra {\bf B} is generated by $\bar
b,b$
and H and endowed with the folowing Hopf algebraic structure}
$$\Delta(x)=x\otimes 1+ 1\otimes x,$$
$$S(x)=-x,\epsilon(x)=0,x=\bar b,b$$
$$\Delta(H)=H\otimes 1+1\otimes H +\bar b \otimes b\eqno{(2.7)}$$
\vspace{0.6cm}

{\bf 3.Quatum Double and Universal R-matrix}
\vspace{0.4cm}

It should be noticed that the dual Hopf algebraic structure of {\bf B}
can also be obtained from the formal group theory [7] of Lie algebra in
principle where but the explicit expressions of the e Baker-Comppell-Hausdorff
formula  for the HW Lie algebra .In this sense the Drinfeld's theory is not
the unique approach to get the dual Hopf algebraic structure.However ,it is
important that  Drinfeld's theory can also provide us with a convenient
method to `combine' {\bf A} and {\bf B} to form a `larger' Hopf algebra
{\bf D} containing {\bf A} and {\bf B} as subalgebras.The universal R-matrix
for QYBE can be automatically given in this construction.

According to  the multipliction formula for the quantum double
$$ba=\sum_{i,j} <a_i(1),S(b_j(1))><a_i(3),b_j(3)>a_i(2)b_j(2) \eqno{(3.1)}$$
where $c_i(k)(k=1,2,3;c=a,b)$ are defined by
$$\Delta^2(c)=(id\otimes\Delta)\Delta(c)=(\Delta\otimes id)\Delta(c)=
\sum_i c_i(1)\otimes c_i(2)\otimes c_i(3)$$
and we use the explicit expressions
$$\Delta^2(H)=H\otimes 1\otimes 1+1\otimes H\otimes 1+1\otimes1\otimes H +
1\otimes \bar b \otimes b+\bar b\otimes b\otimes 1+\bar b\otimes 1\otimes b
\eqno{(3.2)}$$
to prove the following results

{\bf Proposition 3.}{\it
The quantum double {\bf D} is generated by $a,\bar a,b,
\bar b,E,H$ as an associative algebra with the only nonzero commutators}
$$[a,\bar a]=E,[H,a]=\bar b,[H,\bar a]=-b,
\eqno{(3.3)}$$
{\it and as a non-cocomutative Hopf algebra with the structure (2.2) and (2.7).
The universal R-matrix ,a canonical element intwining {\bf A } and {\bf B},is}
$${\it R}=\sum^{\infty}_{m,n,s=0}X(m,n,s)\otimes Y(m,n,s)$$
$$=exp(\bar a\otimes \bar b)exp(a\otimes b)exp(E\otimes H), \eqno{(3.4)}$$

Using the above commutation relations,we can directly verify the the following
quasi-triangular relations

$$\hat{R}\Delta(x)=\sigma\Delta(x)\hat{R},$$
$$(\Delta\otimes id)\hat{R}=\hat{R}_{13}\hat{R}_{23},$$
$$(id\otimes\Delta)\hat{R}=\hat{R}_{13}\hat{R}_{12}, \eqno{(3.5)}  $$
$$(\epsilon\otimes id)\hat{R}=1=(id\otimes\epsilon)\hat{R},$$
$$(S\otimes id)\hat{R}=\hat{R}^{-1}=(id\otimes S)\hat{R},$$
where $\sigma$ is such a permutation that $\sigma(x\otimes y)=y\otimes x
,x,y\in D.$The eqs.(3.5) imply that the above constructed universal R-matrix
satisfies the abstract QYBE
$$\hat{R}_{12}\hat{R}_{13}\hat{R}_{23}=\hat{R}_{23}\hat{R}_{13}\hat{R}_{12},
\eqno{(3.6)}$$
where $\hat{R}_{12}=\sum_m a_m\otimes b_m\otimes 1,
\hat{R}_{13}=\sum_m a_m\otimes 1 \otimes b_m,
\hat{R}_{23}=\sum_m 1\otimes a_m\otimes b_m$
and  $a_m$ and $b_m$ are the dual bases vectors of {\bf A} and {\bf B }
respectively.Here,we simply note $\hat{R}=\sum_m a_m\otimes b_m$.
\vspace{0.5cm}

{\bf 4. On Representations and Realizations of the Quantum Double}
\vspace{0.4cm}

In order to obtain the R-matrices for QYBE from new universal R-matrix (3.5),
we should consider the representations of the quantum double {\bf D}.For
simplicity we by x denote  $T(x)$ for a representation of {\bf D} as follows.

{\bf Proposition 4}.{\it There does not exist a finite dimensional
irreducible representation of {\bf D} besides the trivial representations
T for which there at least is one generator s such that T(x)=0.}

{\bf Proof}.Thanks to the Schur lemma we know that the representatives of
the central elements  $E,{\bar b}$ and b must be nonzero scalars for a finite
dimensional representation ,ie.,
$$E=\eta\neq 0,b=\xi\neq 0,\bar b= \bar\xi\neq 0,\eta,\xi,\bar\xi\in
complex ~field ~{\it C}$$
However,taking the trace of E,we have
$$tr.(E)=tr.([a,\bar a])=0$$
that is $\eta = 0$.Then,a contrdiction appears.

we learen from this proposition and its proof that the finite dimensional
representation of {\bf D} must be neither irreducible nor a sum of some
non-triviallyirreducible representations.The possible non-trivial
finite dimensional representations are only those indecomposable ones,
the  reducible but not completely reducible representations where $tr.(E)=0.$
For the latter we can give a boson realization
$$a=c,\bar a=c^+,b=-\alpha \in {\it C},\bar b=-\beta \in {\it C},$$
$$E=1,H=\alpha c +\beta c^+ , \eqno{(4.1)}
$$
in terms of the boson operators c and $c^+$ satisying
$$[c,c^+]=1,1x=x1=x,x =c,c^+,\eqno{(4.2)}$$
Using the Fork representation of c and $c^+$,
$$c^+\mid n >=(n+1)^{1/2}\mid n +1>,$$
$$c\mid n >=n^{1/2}\mid n-1>,\eqno{(4.3)},$$
on the Fock space
$$\{\mid n >=(n!)^{-1/2}(c^+)^n\mid 0> \mid c\mid 0>=0,n=0,1,2,..\}$$,
we obtain an infinte irreducible representation of
 {\bf D} with explicit matrix  elements
$$(\bar a)_{m,n}=(n+1)^{1/2}\delta _{m,n+1},
(a)_{m,n}=n^{1/2}\delta _{m,n}$$
$$ (E)_{m,n}=\delta _{m,n},(b)_{m,n}=-\alpha\delta _{m,n},
(\bar b)_{m,n}=-\beta\ delta _{m,n}$$
$$(H)_{m,n}=\beta (n+1)^{1/2}\delta _{m,n+1}+n^{1/2}\delta _{m,n}\delta _{m,n
-1},\eqno{(4.4)}$$

In this realization and the corresponding representation,the universal
R-matrix  (3.4) can be expressed as a generator
$${\it R}=e^{-\beta C^+}e^{-\alpha C}\otimes e^{\beta C^+ +\alpha C}
=e^{-\beta\alpha/2}D(-\beta,-\alpha)\otimes D(\beta,\alpha),\eqno{(4.5)}$$
for the two-mode coherent state
$$\mid -\beta,\beta>={\it N}^{-1}\sum ^{\infty}_{m,n=0}\frac{(c^+)^n
\otimes(c^+)^n}{m!n!}\mid 0 >,\eqno{(4.6)}$$
where
$$D(\beta,\alpha)=e^{\beta C^+ +\alpha C}$$
is a non-normalized single coherent state operator.
\vspace{0.6cm}

{\bf 5.Explicit representations}
\vspace{0.4cm}

In this section we consider the explicit construction of  representations
for the quantum double {\bf D} and its universal R-matrix.

Since {\bf D} is the universal enveloping algebra of a Lie algebra with
the basis
$$\{a,\bar a,b,\bar b,H,E \},$$
the PBW theorem determines its basis
$$X[M]=X(m,n,l,r,s,t)=a^m\bar a^nH^lb^r \bar b^sE^t,\eqno{(4.1)}$$
where $m,n,l,r,s,t\in {\bf Z}^+$ and M denotes a 6-vector $M=(m,n,l,r,s, t)$
in a lattice vector space  ${\bf Z}^{+6}$ with the basis
$$e_1 = (1,0,0,0,0,0),e_2 = (0,1,0,0,0,0),$$
$$e_3 = (0,0,1,0,0,0),e_ 4= (0,0,0,1,0,0),$$
$$e_5 = (0,0,0,0,1,0),e_6 = (0,0,0,0,0,1),$$
We can construct an explicit representation of {\bf D} on the basis $X[M]$\\

{\bf Proposition 5}.{\it The regular representation of {\bf D} is}
$$aX[M]=X[M+e_1],$$
$$\bar aX[M]=X[M+e_2]-mX[M-e_1+e_6],$$
$$EX[M]=X[M+e_6],$$
$$bX[M]=X[M+e_4],\eqno{(5.1)}$$
$$\bar bX[M]=X[M+e_5],$$
$$HX[M]=X[M+e_3]+mX[M-e_1+e_5]-nX[M-e_2+e_4],$$
{\bf Proof.}It follows from the following equations
$$[\bar a,a^n]=-nE a^{n-1}$$
$$[a,\bar a^n]=nE\bar a^{n-1},\eqno{(5.2)}$$
$$[H,a^n]=nE\bar b  a^{n-1}$$
$$[H,\bar a^n]=-nEb\bar a^{n-1}$$
which are obtained from  eq.(3.3) by induction.

Let I be a left ideal generated by the element $H-\mu$,ie.,
$$L(\mu)={\bf D}(H-\mu)=\{x(H-\mu)\mid x\in {\bf D}\}$$
Because the left ideal I is a left-invariant {\bf D}-submodule,on the quotient
space $V(\mu)={\bf D}/I(\mu)$:
$$u(K)=a^m\bar a^nb^r \bar b^sE^t Mod.I(\mu)$$
where $ K=(m,n,r,s,t),m,n,r,s,t\in {\bf Z}^+$,the regular representation
induces a infinite dimensional representation

$$au[K]=u[K+e_1],$$
$$\bar au[K]=u[K+e_2]-mu[K-e_1+e_5],$$
$$Eu[K]=u[K+e_5],$$
$$bu[K]=u[K+e_3],\eqno{(5.3)}$$
$$\bar bu[K]=u[K+e_4],$$
$$Hu[K]=\mu u[K]+mu[K-e_1+e_4]-nu[K+e_2+e_3],$$
where
$$e_1 = (1,0,0,0,0),e_2 = (0,1,0,0,0),$$
$$e_3 = (0,0,1,0,0),e_ 4= (0,0,0,1,,0),$$
$$e_5 = (0,0,0,0,1),$$

Now ,let us make a key observation from the eq.(5.3) that the sum
m+n+r+s+t for the basis vectors $u[K]=u(m,n,r,s,t)$ is invariant under
the actions of  {\bf D} .This fact tells us that the following vectors
$$\{u[K]=u(m,n,r,s,t)\mid m+n+r+s+t=N\}$$
for a fixed $N\in {\bf Z}^+$ span an invariant subsapce $V(\mu,N)$ .Then,
the quotient space $Q(\mu,N)=V(\mu)/V(\mu,N)$:
$$Span\{v(K)=u[K]Mod.V(\mu,N)\mid m+n+r+s+t\leq N-1\}$$
is finite dimensional and the dimension is
$$d(N)=\sum ^{N-1}_{k=0}\frac{(k+4)1}{k!4!},\eqno{(5.4)}$$
If we define
$$f_N(K)=\theta(N-1-(m+n+r+s+t))v(k),k=(m,n,r,s,t)$$
where $\theta(x)=1( x\geq 0)$ and  0($x \prec 0$),we can explicitly
write out the above finite dimensional representation in the  explicit
form that is obtained by substituting u[K] in eq.(5.3) by $f_N(K)$.Its
lowest example is a 6-dimensional representation
$$a=E_{1,6},\bar a=-E_{5,1}+E_{2,6},E=E_{5,6},\eqno{(5.5)}$$
$$b=E_{3,6},\bar b=E_{4,6},H=\mu\sum^6_{i=1}E_{i,i}+E_{4,1}-E_{3,2}$$
on an ordered basis
$$f_1(1,0,0,0,0),f_1(0,1,0,0,0),f_1(0,0,1,0,0)$$
$$f_1(0,0,0,1,0),f_1(0,0,0,0,1),f_1(0,0,0,0,0)$$
where $E_{i,j}$ are the materix units with the corresponding elemets
$$(E_{i,j})_{r,s}=\delta_{i,r}\delta_{j,s}$$

One purpose of building  quantum double is to obtain the solutions
of the QYBE in terms of its universal R-matrix and matrix representations.
In order to find the solutions  of QYBE associated with the
exotic quantum double D,we have studied  representation theory and construct
both finite and infinite dimensional representations of {\bf D}.In fact,
for a given representation $T^{[x]}$ of {\bf D}:
$$T^{[x]} : {\bf D}\rightarrow End(V)$$
on the linear space V where x is a continuous parameter,we can construct
a R-matrix
$$ R(x,y)=T^{[x]}\otimes  T^{[y]} (\hat{R})$$
satisfying the QYBE
$$R_{1,2}(x,y)R_{1,3}(x,z)R_{2,3}(y,z)=R_{2,3}(y,z)R_{1,3}(x,z)R_{1,2}(x,y),
\eqno{(5.6)}$$
Here,x,y and  z appear as the color parameters [24] similar to the
non-additive spectrum parameters in QYBE.For example, using the above
obtained 6-dimensional representation,we can construct a $36\times 36$-
R-matrix
$$R=exp(E_{1,6}\otimes E_{3,6})exp([-E_{5,1}+E_{2,6}]\otimes E_{4,6} )
exp(E_{5,6} \otimes(\mu\sum^6_{i=1}E_{i,i}+E_{4,1}-E_{3,2}))$$
$$=(1+E_{1,6}\otimes E_{3,6})(1+[-E_{5,1}+E_{2,6}])
(1+E_{5,6} \otimes(\mu\sum^6_{i=1}E_{i,i}+E_{4,1}-E_{3,2}),\eqno{(5.7)}$$

It is need to pointed out that the higher dimensional representations
can also be obtained in the same form.
\vspace{0.5cm}

{\bf 6.Discussions}
\vspace{0.7cm}

To conclude this paper,we should give some remarks on our exotic quatum
double and its relations to the known results.

 From the construction of the exotic quantum double in this paper,
we can see that a commutative(Abelian) algebra ,eg.,the subalgebra {\bf B},
can be
endowed with a non-cocommutative Hopf algebraic structure and its quatum
daul {\bf A} and quantum double {\bf D} can be deduced as  non-commutative
algebras in an inverse process of the construction in this paper.
Such a process  possibly provide
us with a scheme of `quantization' from commutative object to non-commutative
one.An example of this `quantization' was given [5 ] recently.

It has to be pointed out that there are some difficulties in
the futher developments in constructing the general quantum double asssociated
with arbitrary Lie algebra.
When one take
the subalgebra B to be the whole UEA of an arbitrary Lie algebra  ,we hardly
write down the dual basis explicitly and so the construction scheme
of this paper can not work well.

In the formal group theory of Lie algbera [7],the bialgebra
structure of the daul to the UEA of a classical Lie algebra can be given
abstractly in terms of the formal group.It is not difficult to further
define  the antipode for this dual bialgebra.So,in this abstract way,
the Hopf algebraic structure can be endowed with to the dual Hopf algebra of
the
 UEA.However,writting out the explicit  Hopf algebraic structure,namely,
the the explicit multiplication relations ,  coproduct , antipode
and  counit for the dual generators,completely depends on the explict
evolution of the Baker-Comppell-Hausdorff formula for classical Lie
algebra.However,it is much difficult to do it even for the simple case
e.g.,SU(2).The study in this paper avoids this evalution so that not
only the dual Hopf algebraic structure is obtained,but also the corresponding
quantum double -the exotic quatum double is built
 for the Borel subalgebra of the UEA of
arbitrary classical Lie algebra by combining the two
subalgebras dual to each other
\vspace{0.5cm}

\large
{\bf Acknowledgements
\vspace{0.5cm}

\large
We wish to express our sincere thanks to Prof.C.N.Yang for drawing our
attentions to the quantum Yang-Baxter equantion and its quantum group theory.
C.P.Sun is supported by Cha Chi Ming fellowship through the CEEC in State
University of New York at Stony Brook.We are also supported in part by the
NFS of China through Northeast Normal University and Nankai Institute of
Mathematics
\newpage

{\bf References\\
\vspace{0.5cm}
\large
\begin{enumerate}
\item L.A.Tankhtajian,{\it Quantum Groups},in Nankai Lect.Series
In Mathematical
Phys.1989,ed by M.L.Ge and B.H.Zhao,World Scientific,1989:\\
M.Jimbo,{\it The Topics from the Representations of $U(g)_q$},in
Nankai Lect.on Math.Phys.ed by M.L.Ge,World Scientific,1992
\item V.G.Drinfeld,Proc.ICM.Berkeley,1986,(ed.By A.Gleason,AMS,1987),p.798;\\
M.Jimbo,Lett.Math.Phys.10(1985)63
\item P.Kulish,N.Y.Reshetikhin,Lett.Math.Phys.18(1989),143.
\item X.F.Liu and C.P.Sun,Science in China A35(1992)73;\\
X.F.Liu and M.L.Ge,Lett.Math.Phys. (1992)197
\item C.P.Sun,X.F.Liu,and M.L.Ge,J.Math.Phys.34(1993),in press\\
C.P.Sun,{\it Canonical Quantization in Terms Quantum Group and
Yang-Baxter Equation },preprint ITP.SB-92-61,1992\\
C.P.Sun,{\it Exotic Quantum Double ,Its Universal R-matrix and
Their representationa,}preprint ITP.SB-92-67,1992\\
\item   L.Biedenharn,J.Phys.A,22(1989),L873;\\
C.P.Sun,H.C.Fu,J.Phys.A,22(1989),L983;\\
A.Macfarlane,J.Phys.A,22(1989),4581.
\item J-P Serre,{\it Lie algebras and Lie Groups },V.A.Benjamin.INY,1965.
\end{enumerate}
\end{document}